\documentstyle[aps,prl,epsfig]{revtex}

\begin{document}

\draft

\title{Entanglement of transverse modes in a pendular cavity}

\author{Stefano Mancini and Alessandra Gatti}

\address{
INFM, Dipartimento di Fisica, Universit\`a di Milano, 
Via Celoria 16, I-20133 Milano, Italy
}

\date{\today}

\maketitle

\widetext

\begin{abstract}
We study the phenomena that arise
in the transverse structure of electromagnetic 
field impinging on a linear Fabry-Perot cavity with
an oscillating end mirror.
We find quantum correlations  
among transverse modes 
which can be considered 
as a signature of their entanglement.
\end{abstract} 


\section{Introduction}

It is now well assessed that an empty optical cavity with a
moving mirror in its steady state may mimic a Kerr medium if 
illuminated with coherent light
\cite{hil84,mey85}. When the mirror is free to oscillate, the 
radiation pressure induces a coupling
between its position and the intensity of the light beam, 
modifying the optical path in an intensity
dependent way. A wide range of applications of this effect 
have been recentely developed \cite{app}.
In particular the system, showing a typical bistable behaviour, 
can be used as a quantum noise eater
device, because the output light is significantly squeezed 
\cite{fab94,mt94}, or to generate highly nonclassical
states for both radiation and mirror \cite{mmt,bose},
due to the nonlinear character of the
interaction. The theoretical treatments of this 
model were fully quantum but at most
one dimensional in space because of the plane wave approximation, 
which ensures that the electric
field is uniform over the transverse plane. 
This means that the investigations dealt only with 
temporal-frequency aspects,
neglecting all features related to space.
On the other hand, 
recent years have seen an increasing interest towards the 
spatial aspects \cite{kol}. 
Hence, the aim of this paper is to study
the spatial phenomena that arise in the transverse structure of 
the light field impinging on a
linear Fabry-Perot cavity with an oscillating end mirror. 
We shall show the possibility 
to correlate transverse modes,
other than to obtain squeezing effects.
Such correlations have different nature, and 
could reveal entanglement 
\cite{sch,ein}.
Moreover, 
we will analyse the differences and
analogies with a Kerr nonlinear system, and we will investigate 
the influence of temperature on
the transverse structures for such optomechanical system.

\section{The Model}

We consider a linear
Fabry-Perot empty cavity with one fixed mirror, partially 
transmitting, and one perfectly
reflecting end mirror. The completely reflecting mirror, 
having a mass $m$, can move, 
back and forth along the cavity axes (say $z$),
undergoing
harmonic oscillations at frequency $\omega_m$.
The amplitude of such oscillations is however, 
much less than the equilibrium cavity length $L$.
The cavity resonances
are calculated in absence of the impinging light. 
The characteristic cavity frequencies are 
assumed many orders of magnitude 
greater than $\omega_m$ to ensure that the number 
of photons generated by the Casimir effect
is completely negligible.
We also assume that the cavity round trip time is much shorter 
than the mirror's period of oscillation and
the Doppler frequency shift of the
photons \cite{unruh} on the moving mirror, 
is completely negligible.

\subsection{The Single Mode Model}

Let us recall the model describing the system in the approximation 
where all the spatial effects (e.g. diffraction) are neglected and the
cavity is assumed to operate in a single longitudinal mode of
frequency $\omega_0$. 
That cavity mode is coherently driven by the action 
of an external field ${\cal E}e^{-i\omega_st}$,
having a frequency $\omega_s$ and a complex amplitude ${\cal E}$.
By indicating with $A$, $A^{\dag}$ the annihiliation and creation operators
of the cavity mode,
the system Hamiltonian, in a frame rotating at the frequency $\omega_s$,
reads \cite{mt94} 
\begin{equation}\label{Hden} 
H=\hbar\Delta
A^{\dag}A+\hbar\omega_m 
\left(\frac{P^2_X}{2}+\frac{X^2}{2}\right)
-\hbar g A^{\dag}A X
+i\hbar\left({\cal E} A^{\dag}-{\cal E}^* A\right)\,, 
\end{equation} 
where $\Delta=\omega_0-\omega_s$ is the cavity detuning;
$X$ and $P_X$ are the dimensionless position and momentum operators of
the mirror, obeying the commutation relation $[X,P_X]=i$. 
The interaction part (third term) of Eq.(\ref{Hden}) accounts for the
effect of radiation pressure  
force which causes the instantaneous displacement of the mirror, 
and the coupling constant is given by \cite{mt94}
\begin{equation}\label{coup}
g=\frac{\omega_0}{L}\sqrt{\frac{\hbar}{m\omega_m}}\,.
\end{equation} 

In writing down the equations describing the dynamics of the system, 
we must take into account
the damping of the movable mirror due to the coupling with a thermal 
bath in equilibrium at
temperature $T$, and the cavity losses, due to the coupling of the 
internal mode with all the
external modes of radiation through the fixed  (transmitting) mirror. 
Hence, we have the following quantum Langevin equations 
\begin{eqnarray}
\partial_t\,A(t)&=&-i \Delta A(t)+i g X(t) A(t) +{\cal E}
- \gamma_c\,A(t)
+\sqrt{2\gamma_c} \,A_{\rm in}(t)\,,
\label{lanAsingle}\\
\partial_t\,X(t)&=&\omega_m P_X(t)\,,
\label{lanXsingle}\\
\partial_t\,P_X(t)&=&-\omega_m X(t)+g A^{\dag}(t) A(t) 
- \gamma_m P_X(t)
+\sqrt{\gamma_m} \,\epsilon_{\rm in}(t)\,,
\label{lanPsingle}
\end{eqnarray}
where $\gamma_c$, $\gamma_m$ are the
decay rates of the cavity mode and of the mirror momentum, 
respectively. 
The noise operators (labeled with the subscript ``in'') have zero 
expectation value, 
and obey the
following  correlations:
\begin{equation}\label{Acorr}
\langle A_{\rm in}(t)A_{\rm in}(t')\rangle=0\,,
\quad
\langle A_{\rm in}(t)A^{\dag}_{\rm in}(t')\rangle=\delta(t-t')\,,
\end{equation}
\begin{equation}\label{epscorr}
\langle \epsilon_{\rm in}(t)\epsilon_{\rm in}(t')\rangle=
N_T\delta(t-t')\,,
\end{equation}
where $N_T$ is the number of thermal excitations of the mirror
$N_T=k_BT/\hbar\omega_m$,
with $k_B$ being the Boltzmann constant.
The form of Eq.(\ref{lanAsingle}) with correlations (\ref{Acorr}) 
corresponds to the quantum optical master equation \cite{gard}.
Instead, the form of Eqs.(\ref{lanXsingle}), (\ref{lanPsingle}), 
with correlations (\ref{epscorr}), 
corresponds to the standard quantum Brownian master equation \cite{gard}.
The latter is only valid in the limit $N_T \gg 1$, while a more careful
analysis is required in the opposite limit \cite{lowT}.

\subsection{The Spatially Multimode Model}

Motivated by the recent progress in the study of transverse quantum 
effects in optical systems \cite{kol} we wish to extend the 
previous model 
to the case of spatially multimode fields.
Let us again assume the validity of the single longitudinal mode
approximation, which, toghether with the mean field approximation
allows us to neglect the dependence of the field over the 
variable $z$. However, we now allow the radiation field to depend on
the transverse vector ${\bf x}\equiv (x,y)$,
which is the position vector in the plane orthogonal 
to the direction $z$ of propagation of fields. 

Then,  Eqs.(\ref{lanAsingle}), (\ref{lanXsingle}), 
(\ref{lanPsingle}) after elimination of $P_X$, must be rewritten as
\begin{eqnarray}
\partial_t\,A({\bf x},t)&=&-i \Delta A({\bf x},t)
+i g X(t) A({\bf x},t) +{\cal E}({\bf x})
-\gamma_c \,A({\bf x},t)
+i\gamma_c \ell_D^2 \nabla_{\perp}^2 A({\bf x},t)
+\sqrt{2\gamma_c} \,A_{\rm in}({\bf x},t)\,,
\label{lanA}\\
\partial_t^2\,X(t)&=&-\omega_m^2 X(t)
+g\omega_m \int \,d{\bf x}\, A^{\dag}({\bf x},t) A({\bf x},t) 
-\gamma_m \partial_t X(t)
+\omega_m \sqrt{\gamma_m} \,\epsilon_{\rm in}(t)\,,
\label{lanX}
\end{eqnarray}
where the transverse Laplacian 
$\nabla^2_{\perp}\equiv\partial^2/\partial x^2+
\partial^2/\partial y^2$ 
has been introduced to
describe the diffraction in the paraxial approximation \cite{bw}.
The coefficient $\ell_D^2$ has the dimension of an area, and is given by 
$\ell_D^2=c^2/2\omega_s\gamma_c$. It represents the 
typical length scale for 
spatial structures emerging in an optical resonators.
In the following we shall rescale all the spatial lengths over $\ell_D$. 

For the sake of simplicity we considered only the case of 
a uniform mirror motion over the transverse plane, 
even if there is the possibility to take into account
acoustic modes of the resonators \cite{heid}, 
but this is planned for a future work.

In order to avoid difficulties arising from the continuum 
of transverse modes, we consider in
the transverse plane $(x,y)$ a square of side $\ell$, and 
we assume periodic boundary conditions for
the fields. A complete set of transverse modes 
(corresponding to a single longitudinal resonance)
is then given by
\begin{equation}\label{set}
f_{\bf n}({\bf x})=\frac{1}{\ell}
\exp\left(i{\bf k}_{\bf n}\cdot{\bf x}\right)\,,
\quad
{\rm with}
\quad
{\bf k}_{\bf n}=\frac{2\pi}{\ell}{\bf n}\,,
\quad
{\bf n}\equiv(n_x,n_y)\,,
\end{equation}  
where $(n_x,n_y)$ is a couple of integer numbers, 
$n_x,n_y=0,\pm 1,\pm 2\ldots$.
Then, the fields can be expanded in the following way 
\begin{eqnarray} 
A({\bf x},t)&=&\sum_{{\bf n}}f_{\bf n}({\bf x})a_{{\bf n}}(t)
=\frac{1}{\ell}\sum_{{\bf n}}e^{i{\bf k}_{\bf n}\cdot{\bf
x}}a_{{\bf n}}(t) \,,
\label{Aexp}\\ 
{\cal E}({\bf x})&=&\sum_{{\bf n}}f_{\bf n}({\bf x})e_{{\bf n}}
=\frac{1}{\ell}\sum_{{\bf n}}
e^{i{\bf k}_{\bf n}\cdot{\bf 
x}}e_{{\bf n}}\,.
\label{Eexp} 
\end{eqnarray} 
The Hamiltonian (\ref{Hden}) of the single mode model 
can thus be generalized as
\begin{equation}\label{H} 
H=\hbar\omega_m \left(\frac{P^2_X}{2}+\frac{X^2}{2}\right)
+\sum_{{\bf n}}\left[\hbar\Delta_{{\bf n}}
a^{\dag}_{{\bf n}}a_{{\bf n}}
\right]
-\hbar g\sum_{{\bf n}}\left[a^{\dag}_{{\bf n}}a_{{\bf n}}
X\right]
+i\hbar\sum_{{\bf n}}\left[a^{\dag}_{{\bf n}}e_{{\bf n}}
-a_{{\bf n}}e^*_{{\bf n}}\right]\,,
\end{equation}
where we introduced the mode detuning $\Delta_{\bf n}
=\omega_{\bf n}-\omega_s$, with
\begin{equation}\label{omegan}
\omega_{{\bf n}}=\omega_0
+\gamma_c {\bf k}^2_{{\bf n}}\,,
\end{equation}
the frequency of the transverse mode ${\bf n}$.

We are now able to derive from Eq.(\ref{H})
the Langevin equation for each transverse mode
\begin{eqnarray}
\partial_t \,a_{\bf n}(t)&=&
-i\Delta_{\bf n}a_{\bf n}(t)+i g X(t) a_{\bf n}(t)
+e_{\bf n}(t)
-a_{\bf n}(t)+\sqrt{2}\, 
a_{\bf n}^{\rm in}(t)\,,\label{lana}\\
\partial_t^2 \,X(t)&=&-\omega_m^2 X(t)+g\omega_m \sum_{\bf m}
a^{\dag}_{\bf m}(t)a_{\bf m}(t)
-\gamma_m\partial_t X(t)
+\omega_m \sqrt{\gamma_m} \,\epsilon_{\rm in}(t)\,,\label{lanb}
\end{eqnarray}
where we introduced the following scalings
$$
\gamma_c t\to t\,,\;\;
\omega_m/\gamma_c \to \omega_m\,,\;\;
\gamma_m/\gamma_c \to \gamma_m\,,\;\;
g/\gamma_c \to g\,,\;\;
{\cal E}/\gamma_c \to {\cal E}\,,\;\;
\Delta_{\bf n}/\gamma_c\to\Delta_{\bf n}\,,\;\;
A_{\rm in}/\sqrt{\gamma_c}\to A_{\rm in}\,,\;\;
\epsilon_{\rm in}/\sqrt{\gamma_c}\to \epsilon_{\rm in}\,.
$$

Equations (\ref{lana}), (\ref{lanb}) show that each 
mode interacts with the mirror, 
which in turn becomes
an intermediary between the various modes, redistributing the 
quantum information among them.
Depending on such a process we expect that the output light 
will have different characteristics
from the input one.

\section{The Steady State}

We are actually interested in the classical steady state regime  
and in small fluctuations around this steady state. 
To this end, as it is usually in the semiclassical 
treatment of quantum noise,
we set $a_{\bf n}=\alpha_{\bf n}+\delta a_{\bf n}$, $X=x+\delta X$,
where the c-numbers $\alpha_{\bf n}$, $x$ 
are solutions of the classical steady 
state equations
\begin{eqnarray}
0&=&-i\Delta_{{\bf n}}\alpha_{{\bf n}}
+i g x \alpha_{{\bf n}}
+e_{{\bf n}}-\alpha_{{\bf n}}\,,\label{lanass}\\
0&=&-\omega_m^2 x+ g\omega_m \sum_{{\bf m}}
\left|\alpha_{{\bf m}}\right|^2
\,,\label{lanxss}
\end{eqnarray}

By eliminating the variable $x$ from 
Eqs.(\ref{lanass}), (\ref{lanxss}), 
it is possible to get an infinite set of 
coupled cubic equations
\begin{equation}\label{sseq}
e_{\bf n}=
\alpha_{\bf n}
\left\{1+i\left[
\sum_{\bf m}\left|\alpha_{\bf m}\right|^2
-\Delta_{\bf n}\right]\right\}\,,
\end{equation}
where we have scaled the variables as follows
\begin{equation}\label{scalP}
\left(\frac{g^2}{\omega_m}\right)^{1/2}\alpha_{\bf n}
\to \alpha_{\bf n}\,,\quad
\left(\frac{g^2}{\omega_m}\right)^{1/2}e_{\bf n}
\to e_{\bf n}\,.
\end{equation}
In the case of only one spatial mode, 
the above set of equations reduces 
to only one cubic equation which
shows a typical bistable behavior \cite{mt94}.

To proceede on we consider, without loss of generality, 
a gaussian and real pump
\begin{equation}\label{gaupump}
{\cal E}({\bf x}\,)=\sqrt{\frac{2}{\pi}}
\frac{{\cal E}}{w_p}
e^{-(x^2+y^2)/w_p^2}\,,
\end{equation}
where $w_p$ idicates the pump waist
(we will always use $\ell \gg w_p$). 
The total input power is given by
\begin{equation}\label{Pin}
P_{\rm in}={\cal E}^2=\int\,d{\bf x}\,
|{\cal E}({\bf x}\,)|^2\,,
\end{equation}
while the pump components on the spatial modes are
\begin{equation}\label{en}
e_{\bf n}={\cal E}\frac{\sqrt{2\pi}w_p}{\ell}
\exp\left[
-\left(\frac{w_p}{2}\right)^2 {\bf k}_{\bf n}^2
\right]\,.
\end{equation}

Now, by defining the intracavity power $P$ as
\begin{equation}\label{P}
P=\int\,d{\bf x}\,
|\alpha({\bf x}\,)|^2
=\sum_{\bf n}|\alpha_{\bf n}|^2\,,
\end{equation}
from Eq. (\ref{sseq}) we can obtain
\begin{equation}\label{sseqms}
|\alpha_{\bf n}|^2=
\frac{|e_{\bf n}|^2}{1+
(P-\Delta_{\bf n})^2}\,,
\end{equation}
and summing over the index ${\bf n}$
\begin{equation}\label{PinP}
P_{\rm in}=P
\left\{
\sum_{\bf n}\frac{
2\pi\left(\frac{w_p}{\ell}\right)^2\exp\left[
-\left(\frac{w_p^2}{2}\right) {\bf k}_{\bf n}^2
\right]}
{1+\left[\Delta_{\bf n}-P\right]^2}
\right\}^{-1}\,.
\end{equation}
This equation gives the functional relation 
between the incident and intracavity intensity.
The bistable behavior for the total intensity is shown in Fig.1.
Depending on the slope of the curve, we have
stable and unstable branches.
Once one has choosen the working point 
along this curve, the input and intracavity
powers are well defined, hence it is possible to 
calculate each single steady state components of the 
intracavity field as
\begin{equation}\label{alP}
\alpha_{\bf n}=\sqrt{P_{\rm in}}\frac{
\sqrt{2\pi}\left(\frac{w_p}{\ell}\right)\exp\left[
-\left(\frac{w_p}{2}\right)^2 {\bf k}_{\bf n}^2
\right]}
{1+i\left[\Delta_{\bf n}-P\right]}\,.
\end{equation}

By considering also the field reflected at the 
transmitting mirror
we have the input-output relation \cite{coll},
in terms of scaled  variables
\begin{equation}\label{alinout}
\alpha_{\rm out}({\bf x})=2\alpha({\bf x})-
\alpha_{\rm in}({\bf x})\,,
\end{equation}
where ${\cal E}({\bf x})\equiv\alpha_{\rm in}({\bf x})$.
It is possible to see, in Fig. 2,
that the Gaussian profile
of the beam is maintained from input to
outgoing fields.

Finally, the stationary displacement of the mirror due
to the radiation pressure results $x=P/g$.

\section{Dynamics of Small Fluctuations}

The evolution equations for small fluctuations coming 
from the linearization of the 
Eqs.(\ref{lana}), (\ref{lanb})
around a stable steady state are
\begin{eqnarray}
\partial_t\, \delta a_{\bf n}(t)&=&
-i\Delta_{\bf n}\delta a_{\bf n}(t)
+i \sqrt{\omega_m} \delta X(t)
\alpha_{\bf n}
+i P \, \delta a_{\bf n}(t)
-\delta a_{\bf n}(t)
+\sqrt{2}\, \delta a_{\bf n}^{\rm in}(t)\,,\label{lanflua}\\
\partial_t^2 \,\delta X(t)&=&-\omega_m^2 \delta X(t)
+\omega_m\sqrt{\omega_m} \sum_{\bf m}
\left[\alpha_{\bf m}\delta a^{\dag}_{\bf m}(t)
+\alpha^*_{\bf m}\delta a_{\bf m}(t)\right]
-\gamma_m\partial_t\delta X(t)
+\omega_m \sqrt{\gamma_m}\, \epsilon_{\rm in}(t)\,.\label{lanfluX}
\end{eqnarray}
We immediately recognize that in the case of flat pump, 
i.e. $\alpha_{{\bf n}}=\alpha_{{\bf
0}}\,\delta_{{\bf n},{\bf 0}}$, the coupling with the mirror 
occours only for one mode (the
fundamental), hence no transverse effects can arise. This is 
one of the peculiar differences with
respect to the Kerr-like model, and it is due to the fact that 
only one mode practically survive at
steady state.

In the frequency domain Eqs.(\ref{lanflua}), 
(\ref{lanfluX}) become
\begin{eqnarray}\label{lanflucOm}
i\Omega\,\delta {\tilde a}_{{\bf n}}(\Omega)&=&
-i\Delta_{{\bf n}}\delta {\tilde a}_{{\bf n}}(\Omega)
+i \sqrt{\omega_m} \delta {\tilde X}(\Omega)
\alpha_{{\bf n}}
+i P \, \delta {\tilde a}_{{\bf n}}(\Omega)
-\delta {\tilde a}_{{\bf n}}(\Omega)+\sqrt{2} \,
\delta {\tilde a}_{{\bf n}}^{{\rm in}}(\Omega)\,,\\
-\Omega^2\,\delta {\tilde X}(\Omega)&=&
-\omega_m^2 \delta {\tilde X}(\Omega)
+\omega_m\sqrt{\omega_m} \sum_{{\bf k}}
\left[\alpha_{{\bf k}}\delta {\tilde a}^{\dag}_{{\bf k}}(-\Omega)
+\alpha^*_{{\bf k}}\delta {\tilde a}_{{\bf k}}(\Omega)\right]
-i\Omega\gamma_m\delta {\tilde X}(\Omega)
+\omega_m \sqrt{\gamma_m}\, {\tilde \epsilon}_{\rm in}(\Omega)\,.
\end{eqnarray}

By eliminating the mirror variables one gets an infinite 
set of linear equations
\begin{equation}\label{seteq}
\left[i\Omega
+i\Delta_{{\bf n}}
-i P +1 \right]
\delta {\tilde a}_{{\bf n}}(\Omega)
-i\alpha_{{\bf n}}\chi(\Omega)
\sum_{\bf m}\left[\alpha_{\bf m}
\delta {\tilde a}_{\bf m}^{\dag}(-\Omega)
+\alpha^*_{\bf m}\delta {\tilde a}_{\bf m}(\Omega)\right]
=\sqrt{2}\,\delta {\tilde a}_{{\bf n}}^{{\rm in}}(\Omega)
+i\alpha_{{\bf n}}\,
\chi(\Omega)\sqrt{\frac{\gamma_m}{\omega_m}}\,
{\tilde\epsilon}_{\rm in}(\Omega)\,,
\end{equation}
where we have introduced the mirror response function 
\begin{equation}\label{chi}
\chi(\Omega)=\frac{\omega_m^2}{\omega_m^2
-\Omega^2+i\gamma_m\Omega}\,,
\quad \chi^*(\Omega)=\chi(-\Omega)\,.
\end{equation}

Other typical differences between this model and a Kerr 
nonlinear system are due to the 
dynamics of the moving mirror characterized by a frequency 
dependent susceptibility, and to the presence of a thermal
noise as can be seen from  Eqs.(\ref{chi}) and (\ref{epscorr}).

In Eq.(\ref{seteq}) we may see that the various radiation modes 
can become correlated and some spatial
effects should appear. The latter should also depend 
on temperature.

\section{Solutions}

In order to find the solutions of Eqs.(\ref{seteq}), 
let us write them in a matricial form.
In doing so, we introduce a truncation in the number of modes
effectively achieved, namely we let
$n_x,n_y=-{\overline n},\ldots ,{\overline n}$. 
The truncation is reasonable given the fact that 
the pump field 
supports a finite number of modes.

Then we introduce the following vectors 
\begin{eqnarray}\label{V}
{\cal V}=
\left[
\begin{array}{c}
\left(
\begin{array}{c}
\delta{\tilde a}_{-{\overline n},-{\overline n}}(\Omega)\\
\delta{\tilde a}^{\dag}_{-{\overline n},-{\overline n}}(-\Omega)
\end{array}
\right)
\\
\vdots
\\
\left(
\begin{array}{c}
\delta{\tilde a}_{{\overline n},-{\overline n}}(\Omega)\\
\delta{\tilde a}^{\dag}_{{\overline n},-{\overline n}}(-\Omega)
\end{array}
\right)
\\
\vdots
\\
\vdots
\\
\left(
\begin{array}{c}
\delta{\tilde a}_{-{\overline n},{\overline n}}(\Omega)\\
\delta{\tilde a}^{\dag}_{-{\overline n},{\overline n}}(-\Omega)
\end{array}
\right)
\\
\vdots
\\
\left(
\begin{array}{c}
\delta{\tilde a}_{{\overline n},{\overline n}}(\Omega)\\
\delta{\tilde a}^{\dag}_{{\overline n},{\overline n}}(-\Omega)
\end{array}
\right)
\end{array}
\right]\,,
\quad
{\cal A}=
\left[
\begin{array}{c}
\left(
\begin{array}{c}
i\alpha_{-{\overline n},-{\overline n}}\\
-i\alpha^*_{-{\overline n},-{\overline n}}
\end{array}
\right)
\\
\vdots
\\
\left(
\begin{array}{c}
i\alpha_{{\overline n},-{\overline n}}\\
-i\alpha^*_{{\overline n},-{\overline n}}
\end{array}
\right)
\\
\vdots
\\
\vdots
\\
\left(
\begin{array}{c}
i\alpha_{-{\overline n},{\overline n}}\\
-i\alpha^*_{-{\overline n},{\overline n}}
\end{array}
\right)
\\
\vdots
\\
\left(
\begin{array}{c}
i\alpha_{{\overline n},{\overline n}}\\
-i\alpha^*_{{\overline n},{\overline n}}
\end{array}
\right)
\end{array}
\right]\,,
\quad
{\cal D}=
\left[
\begin{array}{c}
\left(
\begin{array}{c}
\Delta_{-{\overline n},-{\overline n}}\\
\Delta_{-{\overline n},-{\overline n}}
\end{array}
\right)
\\
\vdots
\\
\left(
\begin{array}{c}
\Delta_{{\overline n},-{\overline n}}\\
\Delta_{{\overline n},-{\overline n}}
\end{array}
\right)
\\
\vdots
\\
\vdots
\\
\left(
\begin{array}{c}
\Delta_{-{\overline n},{\overline n}}\\
\Delta_{-{\overline n},{\overline n}}
\end{array}
\right)
\\
\vdots
\\
\left(
\begin{array}{c}
\Delta_{{\overline n},{\overline n}}\\
\Delta_{{\overline n},{\overline n}}
\end{array}
\right)
\end{array}
\right]\,.
\end{eqnarray}
They all have $2(2{\overline n}+1)^2$ components.

Now, the system of Eqs.(\ref{seteq}), can be written as
\begin{equation}\label{VinvsV}
\left[{\cal M}+\left(i\Omega+1\right){\cal I}\right]
\cdot{\cal V}=\sqrt{2}\,{\cal V}^{\rm in}
+\sqrt{\frac{\gamma_m}{\omega_m}}\chi(\Omega)
{\cal A}\,{\tilde \epsilon}_{\rm in}(\Omega)\,,
\end{equation}
where the vector ${\cal V}^{\rm in}$ is defined analogously 
to Eq.(\ref{V}), but for the input
operators. Instead, ${\cal I}$ is the 
$2(2{\overline n}+1)^2
\times 2(2{\overline n}+1)^2$ identity matrix, 
while ${\cal M}$ is the 
$2(2{\overline n}+1)^2
\times 2(2{\overline n}+1)^2$ matrix
given by
\begin{equation}\label{M}
{\cal M}_{k,l}=
i(-)^k\left[P
-{\cal D}_k\right]\delta_{k,l}
+i(-)^l\,\chi(\Omega)\,
{\cal A}_k\,{\cal A}_{l-(-1)^l}\,,
\end{equation}

By inverting the relation (\ref{VinvsV}) we can get
the formal solution of the system as
\begin{equation}\label{VvsVin}
{\cal V}={\cal F}\cdot
\left[
\sqrt{2}\,{\cal V}_{\rm in}
+\sqrt{\frac{\gamma_m}{\omega_m}}\chi(\Omega)
{\cal A}\,{\tilde \epsilon}_{\rm in}(\Omega)\right]
\,,
\qquad
{\cal F}=\left[{\cal M}+\left(i\Omega+1\right){\cal I}
\right]^{-1}\,.
\end{equation}

However we are looking for the solution of the outgoing modes.
Then, the input-output relation \cite{coll} can be 
written in vector form as
\begin{equation}\label{inoutvec}
{\cal V}^{\rm out}=\sqrt{2}\,{\cal V}-{\cal V}^{\rm in}\,,
\end{equation}
and if we combine it with Eq.(\ref{VvsVin}), we get
\begin{equation}\label{Vout}
{\cal V}^{\rm out}={\cal B}\cdot{\cal V}^{\rm in}
+{\cal U}{\tilde\epsilon}_{\rm in}\,,
\end{equation}
where
\begin{equation}\label{defB}
{\cal B}=2\,{\cal F}-{\cal I}\,,
\end{equation}
and
\begin{equation}\label{defU}
{\cal U}=\sqrt{2\frac{\gamma_m}{\omega_m}}\,
\chi(\Omega)\,
{\cal F}\cdot{\cal A}\,.
\end{equation}

From the above matrix relations one can extract the expression 
for the various components
$\delta{\tilde a}^{\rm out}_{n_x,n_y}(\Omega)$,
$\delta{\tilde a}^{{\rm out}\,\dag}_{n_x,n_y}(\Omega)$
in terms of the input noise operators.

Finally, it is worth noting that the
transformation among input
and output fields could not preserve
the commutation relations.
This can be understand by observing  
Eq.(\ref{M}) where ${\rm Re}\{{\cal M}_{k,k}\}$
yields and additional damping term $\gamma_{\rm add}
\propto{\rm Im}\{\chi(\Omega)\}$.
Neverthless, if $\gamma_{\rm add} \ll 1$
the commutation relations are preserved.
That happen for istance in the case of $\Omega \to 0$.

\section{The Output Correlations}

Using Eq.(\ref{epscorr}), and the fact that all the elements 
of the vector ${\cal V}^{\rm in}$ are
uncorrelated except those of the form
\begin{equation}\label{Vincor}
\langle{\cal V}^{\rm in}_{2k-1}(\Omega)
{\cal V}^{\rm in}_{2l}(\Omega')\rangle
=\delta\left(\Omega+\Omega'\right)\delta_{k,l}\,,
\end{equation}
we are able to calculate the correlations of the output modes
from the components of Eq.(\ref{Vout}).
They result:
\begin{equation}\label{aodagao}
\langle
\delta{\tilde a}^{{\rm out} \dag}_{\bf n}(-\Omega)
\delta{\tilde a}^{\rm out}_{\bf m}(\Omega')
\rangle
=\Bigg[
\sum_{k=1}^{(2{\overline n}+1)^2}
{\cal B}_{\{{\bf n}\}+2,2k-1}\,
{\cal B}_{\{{\bf n}\}+1,2k}
+N_T\,
{\cal U}_{\{{\bf n}\}+2}\,
{\cal U}_{\{{\bf n}\}+1}
\Bigg]
\delta\left(\Omega+\Omega'\right)\,,
\end{equation}
and
\begin{equation}\label{aoao}
\langle
\delta{\tilde a}^{\rm out}_{\bf n}(\Omega)
\delta{\tilde a}^{\rm out}_{\bf m}(\Omega')
\rangle
=\Bigg[
\sum_{k=1}^{(2{\overline n}+1)^2}
{\cal B}_{\{{\bf n}\}+1,2k-1}\,
{\cal B}_{\{{\bf n}\}+1,2k}
+N_T\,
{\cal U}_{\{{\bf n}\}+1}\,
{\cal U}_{\{{\bf n}\}+1}
\Bigg]
\delta\left(\Omega+\Omega'\right)\,,
\end{equation}
where
$\{{\bf n}\}\equiv 2(2{\overline n}+1)(n_y+{\overline n})
+2(n_x+{\overline n})$.

We now introduce the linearized output intensity operator
\begin{equation}\label{deltaIoutdef}
\delta\,{\tilde I}_{\bf n}^{\rm out}(\Omega)=
\alpha_{\bf n}^{{\rm out}\,*}
\delta\,{\tilde a}_{\bf n}^{\rm out}(\Omega)
+\alpha_{\bf n}^{\rm out}
\delta\,{\tilde a}_{\bf n}^{{\rm out}\,\dag}(-\Omega)\,.
\end{equation}
Then, the spectrum of the output intensity correlations among 
the various modes, namely
\begin{equation}\label{Snm}
S^{\rm out}_{{\bf n},{\bf m}}(\Omega)=
\int\,d\Omega'\;
\frac{
\langle\delta\,{\tilde I}_{\bf n}^{\rm out}(\Omega)
\delta\,{\tilde I}_{\bf m}^{\rm out}(\Omega')\rangle}{
|\alpha_{\bf n}^{\rm out}| |\alpha_{\bf m}^{\rm out}|}\,,
\end{equation}
can be easily calculated by using Eqs.(\ref{aodagao}), (\ref{aoao}).
It should be compared with that of a coherent state,
i.e. $S^{\rm out}_{{\bf n},{\bf m}}(\Omega)=\delta_{{\bf n},{\bf m}}$.

Furthermore, by considering the total output intensity
$\delta {\tilde I}^{\rm out}(\Omega)=\sum_{\bf n}
\delta {\tilde I}^{\rm out}_{\bf n}(\Omega)$,
it is possible to define
the output intensity spectrum
\begin{equation}\label{Sout}
S^{\rm out}(\Omega)=
\int\,d\Omega'\;
\frac{
\langle\delta\,{\tilde I}^{\rm out}(\Omega)
\delta\,{\tilde I}^{\rm out}(\Omega')\rangle}{
\sum_{\bf n}|\alpha_{\bf n}^{\rm out}|^2}\,,
\end{equation}
which, again, can be easily calculated by using Eqs.(\ref{aodagao}), 
(\ref{aoao}).
It should be compared with that of a coehrent state,
i.e. $S^{\rm out}(\Omega)=1$.

\section{Results and Conclusions}

To study numerically the system, 
 we have considered a
pump defined by ($19\times 19$) modes.

In Fig.3 we show the intensity (auto)correlation 
spectrum for each mode, 
$S^{\rm out}_{{\bf n},{\bf n}}(\Omega=0.1)$. 
The squeezing is found
in several modes around the fundamental, 
but it tends to disappear when the temperature 
increases.
The total intensity squeezing $S^{\rm out}(\Omega=0.1)$ has been
calculated for the three cases of Fig.3 obtaining the following
values: 0.11, 0.62, and 1.14. 
The order of magnitude is similar to that obtained in 
Refs.\cite{fab94,mt94},
however, in this case the total squeezing is distributed 
among the various mode.

In Fig.4 we plot the intensity correlation spectrum
of the fundamental mode with each other 
$S^{\rm out}_{{\bf 0},{\bf n}}(\Omega=0.1)$.
At low temperature we have a negative correlation of the
fundamental mode with the neighbourhoods
(the central peak instead represents the autocorrelation of the 
fundamental mode, i.e. its intensity squeezing).
The negative correlations, i.e. anticorrelations, can be understood 
by considering the interaction with the mirror as redistributing 
the photons from
the fundamental (pumped) mode to its neighbourhoods.
On the other hand these anticorrelations could be considered 
as a signature of entanglement.
In fact, thought the notion of entanglement is not clear for
open system, we can see from Eqs.(\ref{Vout}), (\ref{Snm}) 
that the correlations among various modes
arise as consequence of both vacuum noise and 
thermal noise. However, the latter, which is of classsical 
origin, only leads to positive correlations. Instead, the vacuum
noise, purely quantum, can give positive as well as  
negative correlations. 
Therefore, anticorrelations can be considered 
as a signature of purely quantum correlations, hence, entanglement.
Nevertheless, the two types of 
correlations become competing as can be seen
from top to bottom of Fig.\ref{fig4}.
As matter of fact, by increasing the temperature the negative correlations 
tend to disappear,
and the shape tends to assume the form of
$\delta_{{\bf 0},{\bf 0}}$. 
The residual (positive) correlations may be attributed 
to the thermal noise
and could be used to study the Brownian motion of the 
mirror \cite{brow}.
It is to remark, anyway, that the quantum effects are quite
robust to the thermal noise, providing to have an high
quality factor for the moving mirror $\omega_m/\gamma_m$.
This can be easily understood by noticing the factor multiplying
the thermal noise term in the set of equations (\ref{seteq}).

Thought we limited our analysis to the intensity correlations,
we might argue the existence of more fundamental correlations, like
EPR correlations \cite{gmt00}.

In conclusion, we have studied the phenomena related to the finite
extend of a light beam in optomechanical coupling.
In some sense this work can be considered complementary
to Ref. \cite{heid}, where instead several vibrational
modes of the mirror were coupled to only one light mode.
Moreover, the system can give the possibility of 
multimode entanglement which
is one of the most striking aspect of
quantum mechanics \cite{sch,ein}.
Finally, the developed theory could be also used in
gravitational interferometry \cite{abr} where such
transverse effects  could increase the measurement sensitivity.

\bibliographystyle{unsrt}

\begin{figure}[t]
\centerline{\epsfig{figure=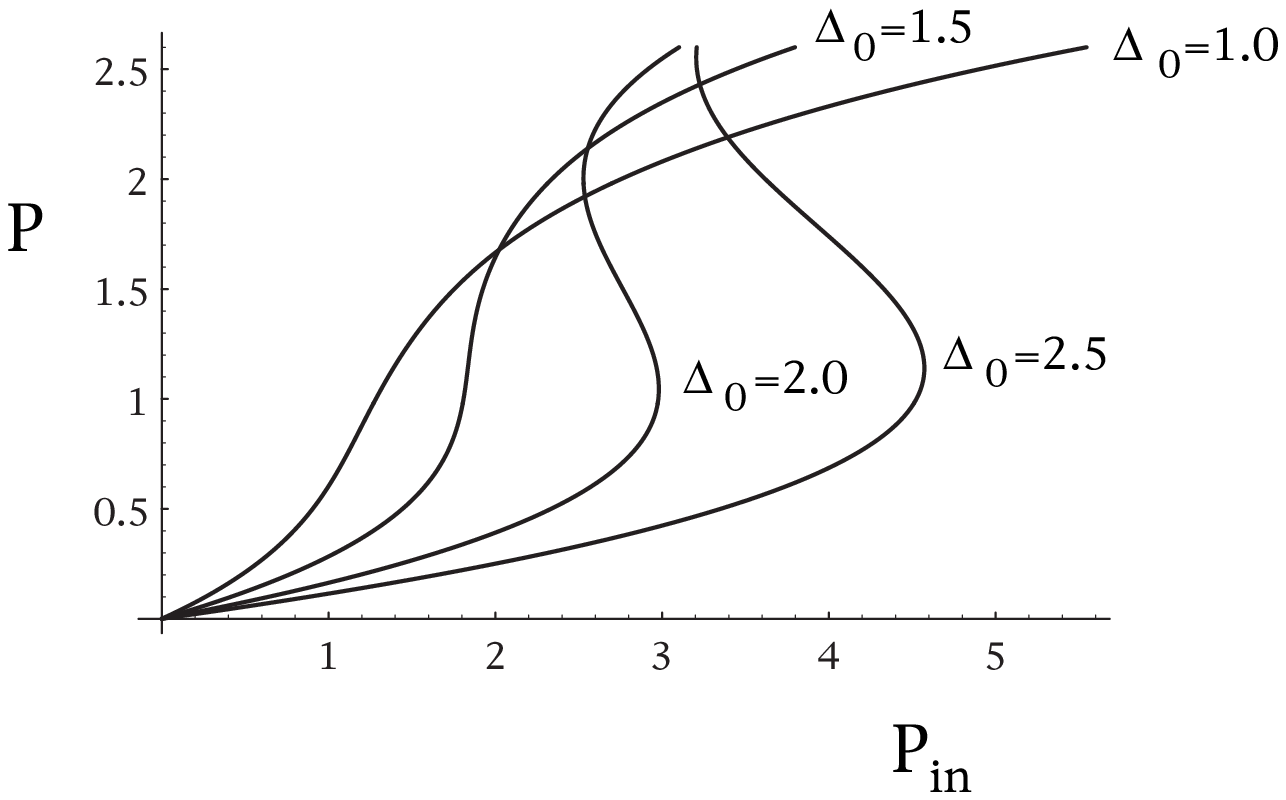,width=3.5in}}
\caption{\widetext 
The bistability curve is represented for several
values of detuning. The chosen pump waist is $w_p=2$.}
\label{fig1}
\end{figure}

\begin{figure}[t]
\centerline{\epsfig{figure=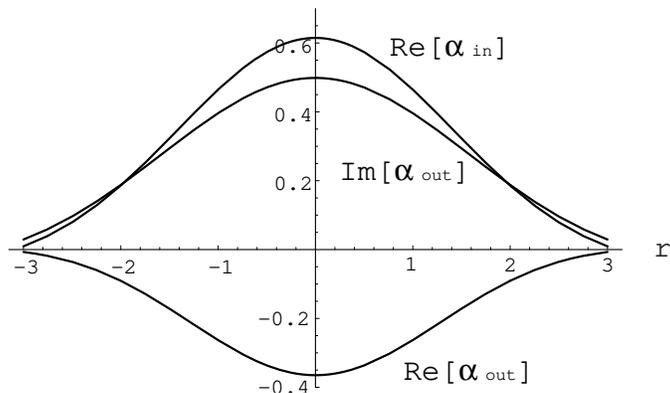,width=3.5in}}
\caption{\widetext
The real and imaginary part
of the input and ouput stationary
fields are represented as a function of $r=(x^2+y^2)^{1/2}$ 
for $w_p=2$, $P_{\rm in}=2.89$, $P=1.06$, $\Delta_{0}=2$.}
\label{fig2}
\end{figure}

\begin{figure}[t]
\centerline{\epsfig{figure=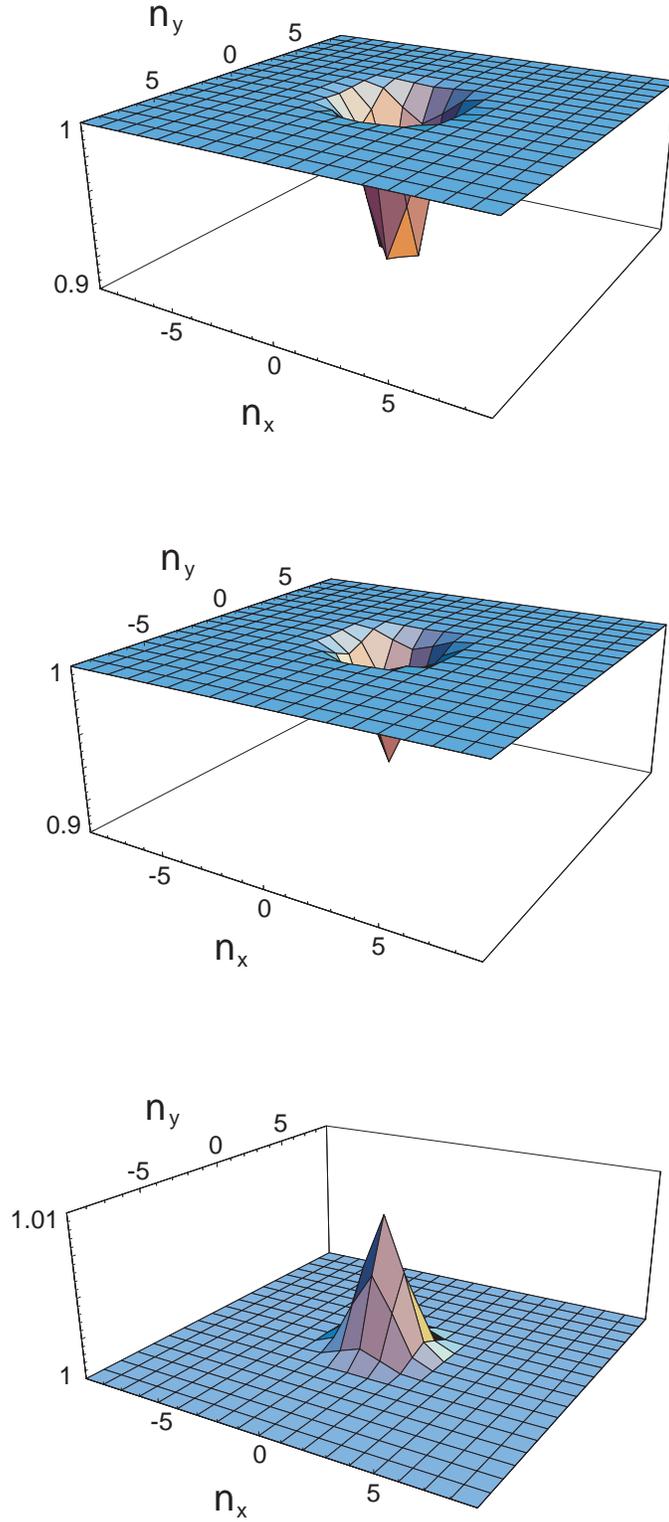,width=3.5in}}
\caption{\widetext
The spectrum $S_{{n},{n}}^{\rm out}(\Omega=0.1)$
is plotted vs ${\bf n}$ for several values of $N_T$
(from top to bottom $N_T=10^4$, $10^5$, $10^6$).
Values of other parameters are: $w_p=2$,
$P_{\rm in}=2.89$, $P=1.06$, $\Delta_{0}=2$,
mechanical quality factor $\omega_m/\gamma_m=10^6$.}
\label{fig3}
\end{figure}

\begin{figure}[t]
\centerline{\epsfig{figure=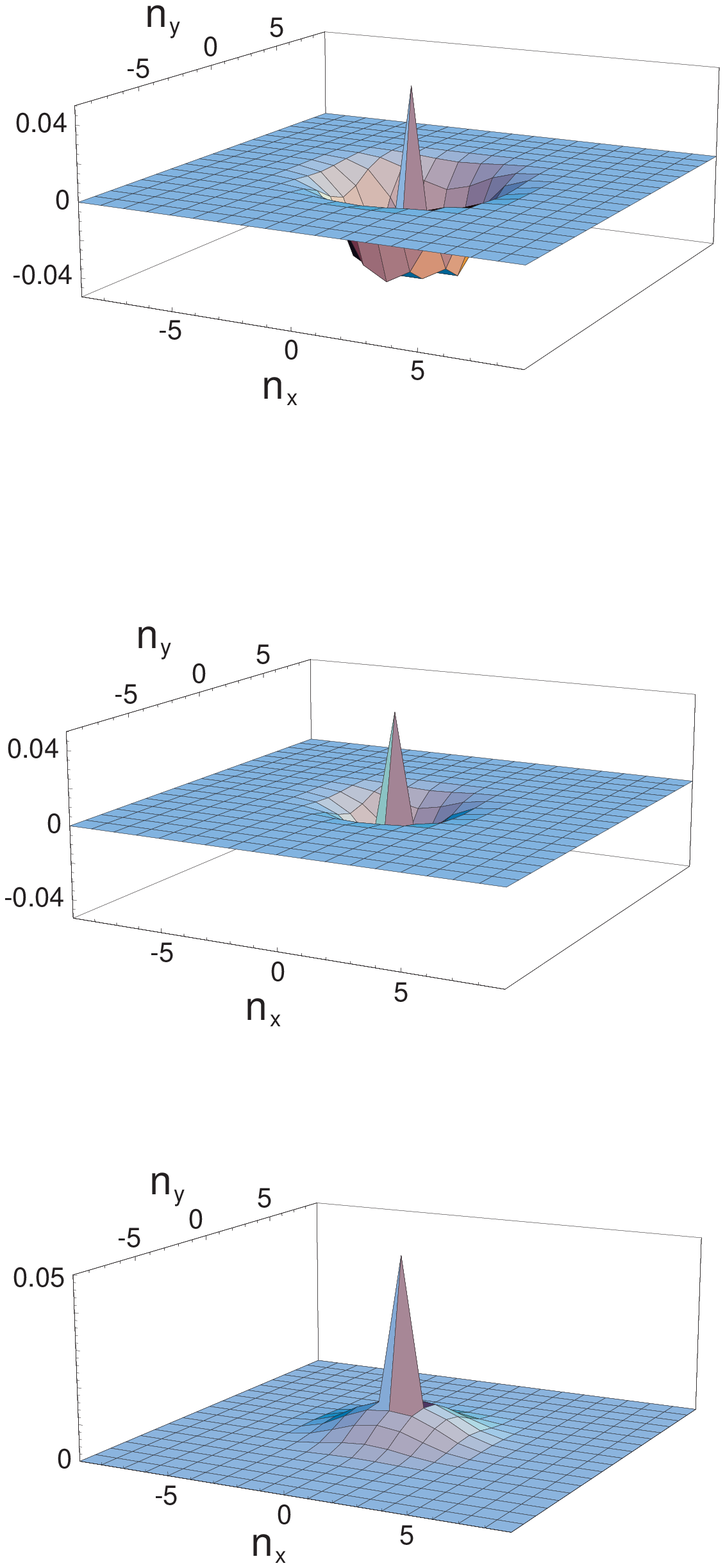,width=3.5in}}
\caption{\widetext
The spectrum $S_{{0},{n}}^{\rm out}(\Omega=0.1)$
is plotted vs ${\bf n}$ for several values of $N_T$
(from top to bottom $N_{T}=10^{4}$, $10^{5}$, $10^{6}$).
Values of other parameters as in Fig.3.
In reality, the central peaks of Fig. 4 go out of the
actual scale, and their values
coincide with those of Fig. 3.}
\label{fig4}
\end{figure}

\end{document}